# Hybrid air-bulk multi-pass cell compressor for high pulse energies with full spatio-temporal characterization

**ALAN OMAR,**[1,*] **MARTIN HOFFMANN,**[1] **GEOFFREY GALLE,**[2] **FRANÇOIS SYLLA**[2] **AND CLARA J. SARACENO**[1]

[1]*Photonics and Ultrafast Laser Science, Ruhr-Universität Bochum, Universitätsstrasse 150, 44801 Bochum, Germany*
[2]*SourceLAB, 7 rue de la Croix Martre, 91120 Palaiseau, France*
*[*alan.omar@ruhr-uni-bochum.de](*alan.omar@ruhr-uni-bochum.de)*

**Abstract:** Multi-pass cell (MPC) compressors have proven to be the method of choice for compression of high average power long-pulse Yb lasers. Yet, generating sub-30 fs pulses at high pulse energy with compact and simple components remains a challenge. This work demonstrates an efficient and cost-effective approach for nonlinear pulse compression at high pulse energy using a hybrid air-bulk MPC. By carefully balancing the relative nonlinear contributions of ambient air and fused silica, we achieve strong spectral broadening without dispersion engineering or pressure-control inside the cell at 400-µJ pulse energy. In this way, we compress pulses from 220 fs to 27 fs at 40.3 W of average power (100 kHz repetition rate), enhancing the peak power from 1.6 GW to 10.2 GW while maintaining 78% of the energy within the main pulse. Our approach combines the strengths of gas-filled and bulk compression schemes and exhibits excellent overall optical transmission (91%) and spectral uniformity. Moreover, we utilize the INSIGHT technique to investigate spatio-temporal couplings and geometrical aberrations of the compressed pulse. Our results demonstrate remarkable temporal homogeneity, with an average Strehl ratio of 0.97 consistently observed throughout the entire spectral profile. Additionally, all spectrally-integrated Zernike coefficients for geometrical aberrations maintain values below 0.02λ.

## 1. Introduction

The pursuit of efficient and versatile pulse compression schemes for high-power Yb-based ultrafast laser systems continues to be a topic of many research efforts aiming to bring these high power systems to the scientific community. In recent years, substantial progress has been made in this direction with the demonstration of the nonlinear pulse compression scheme within Herriott-type multipass cells [1]. These advances have yielded impressive results, reaching high compression ratios while maintaining exceptional overall power efficiency and preserving spatial beam quality across a broad range of pulse energies, spanning from a few microjoules to hundreds of millijoules. This wide range of compression factors has been accomplished by adapting the nonlinear medium within the cell [1–6]. As a result, the landscape of the MPC compression setups can be divided into two primary schemes: bulk-based for moderate energies and gas-filled for higher energies (Fig. 1).

Notably, since the initial demonstration of all-bulk MPCs in 2016 [1], significant milestones have been reached with bulk based MPCs which have been used up to pulse energies of 200 µJ. For instance, in 2022 multi-gigawatt peak power were reported at 200 kHz, compressing 170-µJ pulses from 300 fs to 31 fs with over 88% efficiency, reaching a peak power of 2.5 GW at 200 kHz in a single-stage bulk MPC [7]. Recently, a compact and robust method distributing multiple plates within a single stage showcased its efficacy by compressing 112-µJ pulses from 1.24 ps to 39 fs, yielding an output pulse energy of 65 µJ and a peak power of 1.1 GW [8].

Similarly, using two stages of hybrid multi-plate MPCs achieved burst-mode laser pulse compression down to few cycle pulses. The outcome was the compression of 128.5 µJ pulses from 1.2 ps down to 8.2 fs at an output pulse energy of 56 µJ, corresponding to a peak power of 2.9 GW with relatively low optical transmission of 50% [9]. While the all-bulk MPC compression offers a more compact and cost-effective solution, pulse energies are typically limited to few tens of microjoules due to the high nonlinear refractive indices of bulk media and the correspondingly large B-integral per roundtrip.

On the other hand, while being comparably expensive and bulky, gas-filled MPCs offer a range of distinct advantages [10]. They eliminate the need for dispersion management associated with bulk media, guarantee a more homogeneous distribution and total amount (via pressure) of self-phase modulation (SPM) during propagation, minimize overall losses attributed to optical interfaces, and allow scaling nonlinear temporal compression systems to higher pulse energies. The first gas-filled MPC was demonstrated in 2018 [11]. In this case, an Argon-filled MPC at 7 bar pressure was employed to compress 160-µJ pulses from an initial pulse duration of 275 fs down to 33 fs, corresponding to a 3.25 GW of peak power, while maintaining an overall transmission efficiency of 85%. Since then, multiple reports have been made scaling pulse energy and average power, albeit often at the expense of a rather high system complexity [12,13].

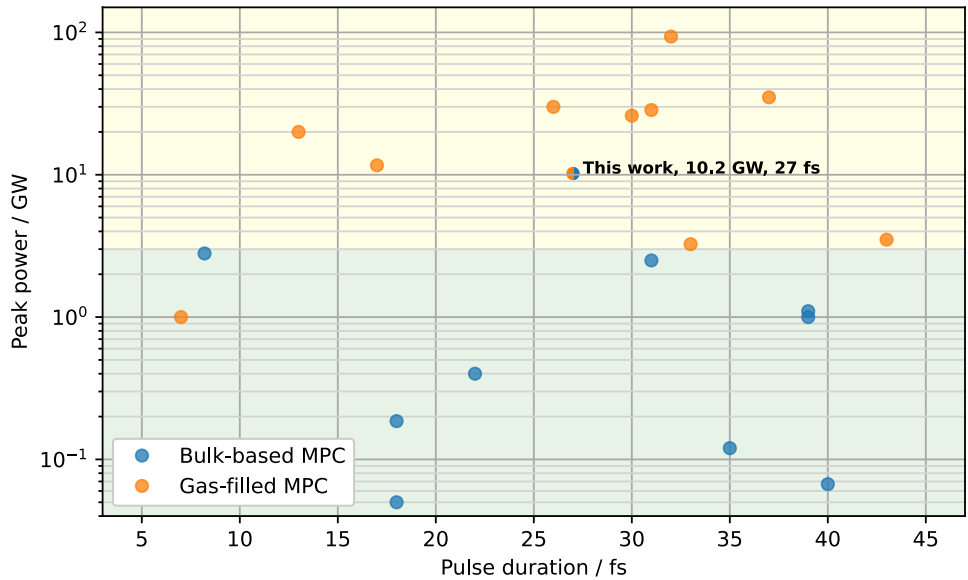

**Fig. 1.** Overview of the MPC compression results at 1030 nm for different types of nonlinear media, pulse duration shorter than 50 fs, and peak power lower than 100 GW.

This work presents a compact and cost-efficient approach for nonlinear pulse compression at high pulse energy and peak power without needing a pressure-controlled cell, i.e. operating in ambient air. We design and demonstrate a hybrid gas/bulk MPC at 400-µJ pulse energy and GW-input peak power, where the gaseous nonlinear medium is ambient air and the bulk medium a plate of fused silica. Both have a significant contribution to the total nonlinearity. The cell employs off-the-shelf quarter-wave high-reflective optics without additional group delay dispersion (GDD). Careful optimization using 3D pulse propagation simulations was required to balance the contribution of both nonlinear media and optimize the MPC's performance. We spectrally broaden 400-µJ, 100-kHz, 220-fs pulses down to 27 fs, effectively reducing the pulse duration by 8.1x and enhancing the peak power from 1.38 GW to 10.2 GW by a factor of 6.4. The resulting pulses are remarkably clean, with 78% of the pulse energy within the main pulse. The overall power efficiency of our setup exceeds 91%. The spectral homogeneity remains uncompromised, with an average spectral overlap of >98%. Additional spatio-temporal measurements demonstrate remarkable temporal homogeneity and exceptional focus quality, attributed to the low aberrations of the compressed beam. To the best of our knowledge, this is the first work of hybrid air-bulk MPC in a single-stage MPC, where such high pulse energies were compressed to sub-30 fs pulses without the need for pressure-controlled MPCs. In fact, by using the air nonlinearity, our cell operates at an order of magnitude higher peak power than state-of-the-art bulk MPC compressors while maintaining the same simplicity advantages. This scheme can operate as a simpler substitute for gas-filled MPCs and fiber compressors within the range of 100 µJs to beyond the mJ level, depending on input pulse duration and constraints on the size of the cell. We believe this opens a new avenue for compact and simple GW-class MPC compressors by merging the advantage of the ambient air and bulk material for simple compression of GW peak power lasers.

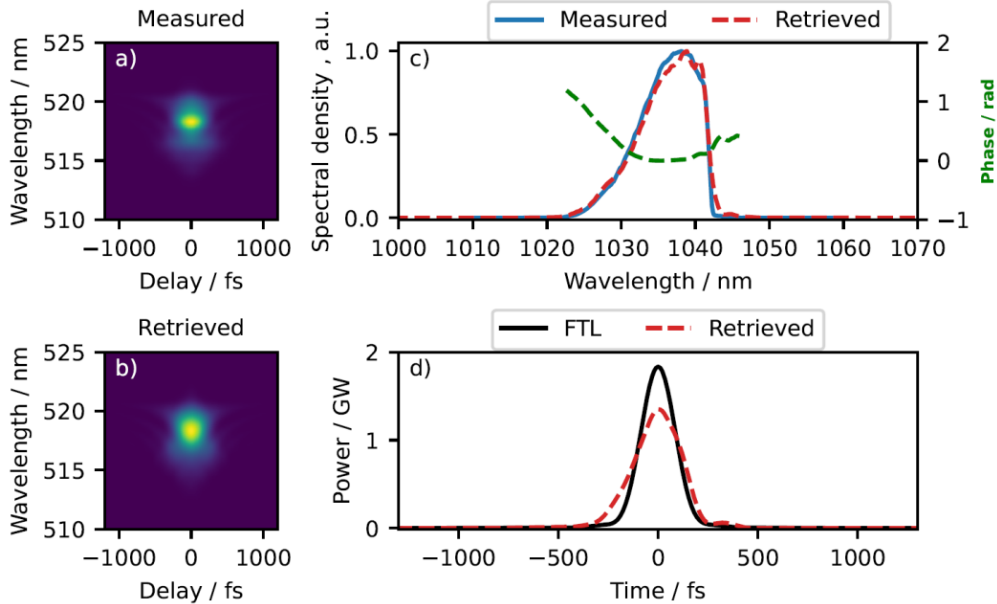

**Fig. 2.** Characterization of the input laser pulse. a) Measured FROG trace. b) Retrieved FROG trace on a grid of 512×512 and an error of 0.38%. c) Measured spectral intensity and retrieved spectrum from FROG. The spectral phase is displayed as a green dashed line indicating the initial chirp. d) Retrieved temporal profile and Fourier transform limit of the measured input spectrum via OSA.

## 2. Laser source

The seed laser is a commercially available Yb-based laser (Light conversion – Carbide) that provides an average power up to 40.3 W and a flexible repetition rate from 100 kHz to 1 MHz. The pulse stretcher enables the pulse duration to be varied from 220 fs to a few picoseconds. In this experiment, the laser was set to the maximum energy of 400 µJ, at a repetition rate of 100 kHz. The laser provides the option to adjust the pulse duration, in our experiment we used this option as a fine tuning knob to obtain best results from the compressor, resulting in most results presented here being achieved with a slightly chirped input pulse (276 fs). The beam quality parameter $M^2$ was ~1.14 in both axes with a slightly asymmetric beam. Fig. 2 shows the characterization of the pulses from the commercial laser system used as input for the MPC using second-harmonic frequency-resolved optical gating (SH-FROG) and an optical spectrum analyzer (OSA). Fig. 2(a) and 2(b) show the measured and retrieved traces, with a retrieval error of 0.38% in a 512×512 grid size. Fig. 2(c) shows the retrieved and measured spectrum, with a full width at half maximum (FWHM) of 9.4 nm, with a cut-off of the spectrum at 1042 nm, supporting a Fourier transform limited (FTL) pulse of 200 fs. Fig. 2 (d) shows the retrieved temporal profile with a FWHM of 276 fs and a peak power of 1.38 GW while the shortest pulse duration available by the laser is 220 fs. In addition, we conducted a spectral homogeneity measurement to investigate the input *V*-parameter, and the intensity-weighted average *V*-parameter is 99.3% and 99.5% for both axes [14].

### 3. Design of MPC

In this study, we employed a numerical model to investigate the hybridization of air and bulk materials as nonlinear media in a single MPC stage for input pulses in a range of a few hundred microjoules to combine the benefits of bulk and gas approaches for spectral broadening in the MPC without the need for pressure-controlled chambers. Our model solves the nonlinear Schrödinger equation in three dimensions (x, y, t) using the split-step Fourier method and considers linear effects such as diffraction, dispersion, and loss, and nonlinear effects such as optical Kerr effect, self-steepening, and instantaneous and delayed Raman scattering [12]. The main criteria for designing the air-bulk MPC are avoiding the ionization of the air at the center of the MPC by managing the peak intensity below 1 TW/cm$^2$ and achieving a significant amount of nonlinearity to enhance the peak power. Woodbury et al [15], have shown that the ionization in the air at atmospheric pressure starts to accumulate at 1 TW/cm$^2$ for 274-fs pulse duration at 1024 nm of wavelength, which is close to the conditions in our work.

We will discuss different cases of a symmetric MPC geometry, where it consists of two identical mirrors with radius of curvature (ROC) *R* and mirror separation *L*. For simplicity, we will consider HR mirrors only, i.e., without a GDD profile. In addition, we constrained ourselves by the available 2-inch mirrors in our labs at the time of the experiment (ROC = 0.5 m, 2 m). Fig. 3(a) shows the beam caustics without nonlinear medium and the peak intensity of one pass through the MPC for different ROCs and cavity lengths for an input pulse energy of 400 µJ and a pulse duration of 276 fs. In Fig. 3(a), the MPC scheme consisted of 0.5-m mirrors, and *L>R* results in a notably higher peak intensity at the focal point. This peak intensity leads to the highest ionization compared to the following scenarios. In the case of *L<R*, we achieve a compact setup, but this also produces a high peak intensity that still causes a weak ionization in the air during the initial passes of the beam. This scheme is suitable for the lower energy case (100 µJ -300 µJ) to achieve a significant amount of nonlinearity, which is out of the scope of this work. Conversely, when R = 2 m and *L<R*, Fig. 3(b) shows that the peak intensity decreases compared to the previous scenarios. This configuration results in minimal changes in beam size throughout the MPC and effectively distributes SPM across the beam's caustic. While *L>R*, this setup allows for higher spectral broadening than the previous scenario (see Fig. 3). Importantly, it still avoids ionization but results in a bulkier MPC, which is impractical for such pulse energy. To conclude this discussion, it was crucial to employ MPC mirrors with larger ROC exceeding 1 m to mitigate ionization and effectively manage the small peak fluence on the mirrors to prevent damage while maintaining a relatively high B-integral for achieving a high compression ratio (see Table. 1).

**Table 1. Advantages and the disadvantages of the hybrid MPC scheme.**

| MPC length *L* | Radius of curvature | |
|---|---|---|
| | 0.5 m | 2 m |
| *L<R* | − Low nonlinearity<br>− ionization | + high nonlinearity<br>+ no ionization |
| *L>R* | + high nonlinearity<br>− ionization | + high nonlinearity<br>+ no ionization<br>− bulkiness |

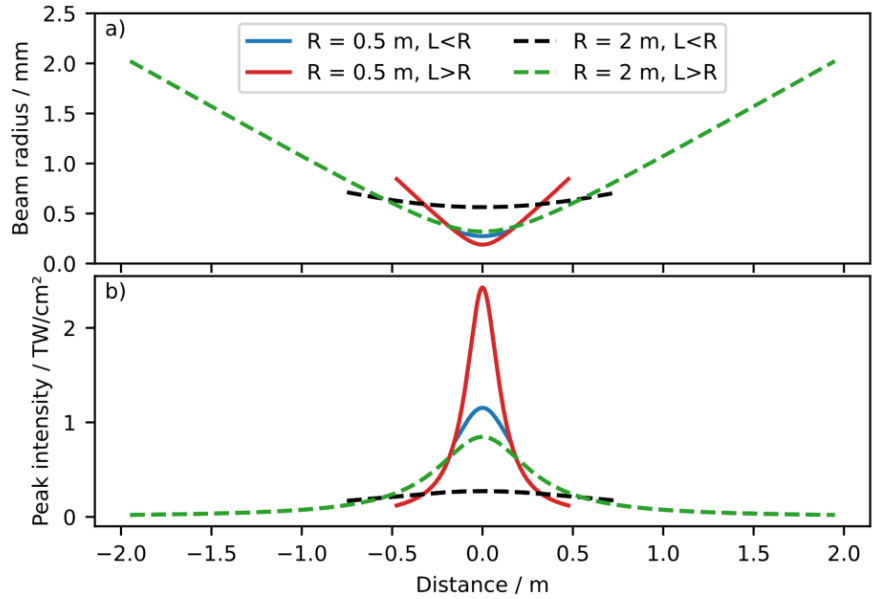

**Fig. 3.** The beam caustic without nonlinear media and the peak intensity of one pass through the MPC. a) the beam caustic of standard MPC of two identical mirrors with ROC $R$ and mirror separation L with 403 µJ and pulse duration of 276 fs. b) the corresponding peak intensity profile of the ideal Gaussian temporal profile across the previous MPC schemes.

We investigated the spectral broadening and compressibility of the L<R case, and we constrained ourselves to 135 cm MPC length, which is the maximum allowed by size constraints of the optical setup. The highest and practical number of passes allowed by this MPC separation was 46 passes through the nonlinear medium i.e., 23 spots on 50-mm MPC mirrors without clipping on the in-coupling mirror with 1.4-mm beam size on the mirror. Our simulations encompassed two distinct scenarios. In the first scenario, we employed HR mirrors only, i.e., the MPC operates in a positive chirp regime due to the material dispersion in addition to the pre-chirp of the input pulse. In the second scenario, the mirrors have a GDD profile to compensate for the material dispersion of the FS and the air. We initiated the simulation with 400 µJ pulse energy and chirped pulse duration of 276 fs with fundamental transverse mode $TEM_{00}$ i.e., with ideal Gaussian beam profile. Fig. 4(a) shows the spectral broadening results of the hybrid air-bulk MPC for both study cases. In the case of dispersive mirrors, the overall B-integral is 21.6 rad (0.47 rad per pass), achieving spectral broadening supports FTL of 18 fs with FS contribution of ~35.7% to the overall B-integral with 7.6 rad. While the overall B-integral in the case of HR only is 16.56 rad (0.36 rad per pass), supporting an FTL of 25 fs and the FS plate contribution is ~35.6% to the overall B-integral with 5.9 rad. In this case, the B-integral per pass and the nonlinearity overall decrease continuously in the cell because of the gradual decrease in peak power due to dispersive temporal pulse-broadening in the FS plate (Fig 4(c)). Moreover, the oscillation of the B-integral with the roundtrips is due to the nonlinearity-related change of the beam caustics from roundtrips to another.

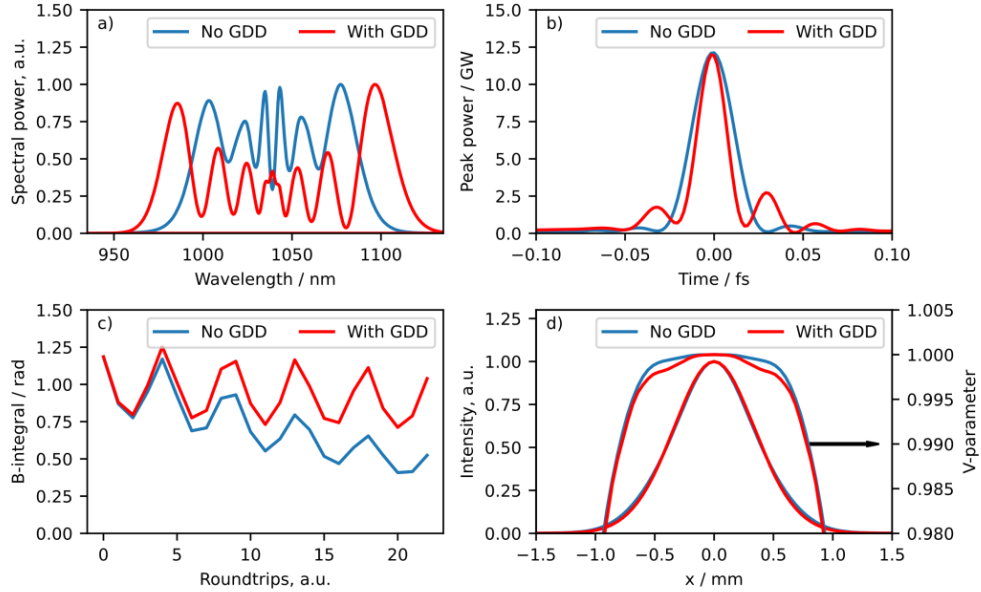

**Fig. 4.** Simulation results of hybrid air-bulk MPC with input pulse energy of 403 µJ and 276 fs pulse duration for study cases: with GDD profile on MPC's mirrors to compensate material dispersion and without GDD. a) Simulated spectral broadening in air-bulk MPC. b) The compressed temporal profile showing that both study cases reach the same peak power 12 G. c) The evolution of the B-integral per roundtrips through propagation in the MPC for the study case. d) The spatial profile and the simulated V-parameter after MPC showing high spectral homogeneity V-parameter>99% within the ($1/e^2$) beam area for both cases.

Fig. 4(b) shows the compressed pulse of dispersive mirrors case of 19 fs with a peak power of 12 GW. However, this does not show a peak power enhancement compared to the HR-only case where the pulse is compressed to 27 fs with the same peak power of 12 GW. Thus, for applications where peak power is the main criterion, the cleaner pulse obtained with the same peak power support the idea of utilizing HR mirror only for the MPC and operate the spectral broadening of the MPC in the positive-chirp regime. We note that our main criterion when designing the MPC was to achieve the highest peak power rather than the shortest FWHM pulse duration, as required by our target application in nonlinear conversion to the THz region. The peak fluence on the mirrors is estimated to be 0.05 J/cm$^2$ and well below the HR mirrors' damage threshold. Fig. 4(d) shows the spectral homogeneity of the simulated pulse, with a calculated *V*-parameter >99% in both axes within a $1/e^2$ beam area for both cases. The intensity-weighted average *V*-parameter is ~99.82% for both axes, and the simulated $M^2$<1.05, indicating that this approach is preserving the spectral homogeneity of the beam. Furthermore, a complete spatio-temporal characterization of the pulses is presented in the next section. Fig. 4(d) shows a drop in the *V*-parameter value in the dispersive mirrors case near the wing of the spatial profile, resulting in an intensity-weighted averaged *V*-parameter of ~99.8%. Moreover, in comparison, the net-positive GDD regime approach offers the advantage of designing MPCs for high energies - with peak powers below the critical power of the air - in a more accessible and cost-effective fashion. Furthermore, dispersive mirrors have inferior thermal and damage properties compared with HRs. Lastly, while the net-positive GDD case does indeed have a slightly lower compression efficiency (e.g., spectral bandwidth gain per focus pass is lower), the temporal aspect is in fact identical. In our cell, the peak power is 23% lower than the critical peak power for self-focusing in air [16]. The GDD of the employed thin FS plate at 1030 nm is 19 fs$^2$ per pass, while the GDD of the air per pass is 21 fs$^2$ with an overall material dispersion of 1840 fs² for all passes through the MPC, resulting in an output pulse duration of 616 fs and a peak power of 0.65 GW.

## 4. Experimental setup and compression results

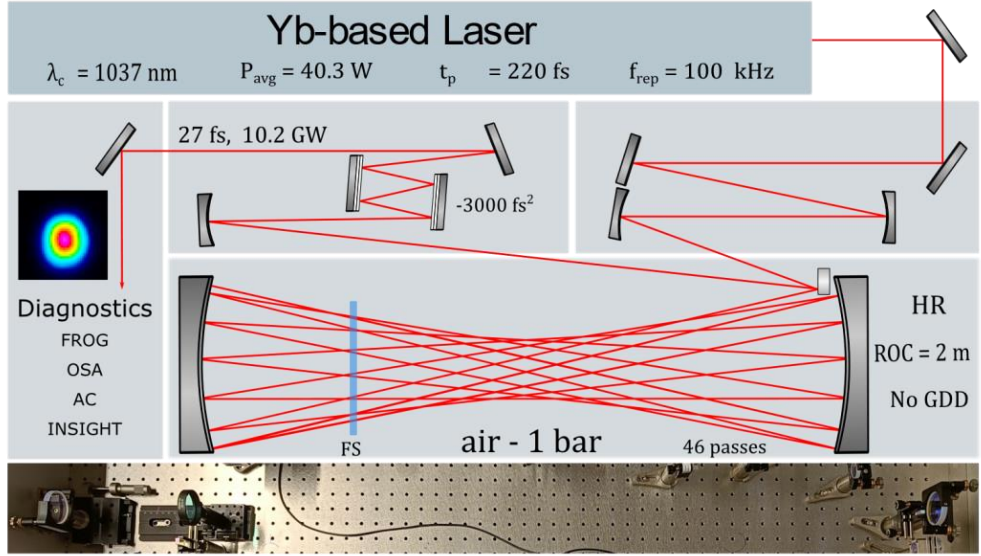

**Fig. 5.** Experimental setup. The seed laser is a commercial laser amplifier, and the Herriott cell mirrors are HR with negligible GDD and separated by 134 cm, providing 46 passes through fused silica. Curved HR mirrors were used to match the mode between the laser and the MPC. The beam is collimated and then compressed with dispersive mirrors with an overall GDD of -3000 fs² at full power, and it is guided for characterization using a wedged sampler.

The experimental setup is depicted in Fig. 5. The output beam of the laser passes through mode-matching optics which consist of two concave mirrors with ROC = 5 m and 12 cm separation. The beam does not go through a focus, and the beam size stays large >2mm therefore the nonlinearity generated in air before the MPC can be neglected. Following these mode-matching optics, the beam enters the MPC in a closed-path configuration via a 5-mm wide rectangular mirror. The hybrid MPC consists of two 2-inch mirrors with a ROC of 2 m each. Both mirrors are highly reflective (HR) with close to zero GDD. The mirrors were separated by 134 cm, providing 23 full roundtrips, corresponding to 46 reflections on either of the MPC mirrors. A 1-mm thick anti-reflection (AR) coated fused silica (FS) plate is placed 20 cm from one mirror in ambient air. At such high peak power levels, it was essential to consider nonlinear mode-matching: considering diffraction alone while ignoring the strong self-focusing effect of the FS plate can lead to the collapse of the beam and result in detrimental peak fluence on the mirrors [17]. However, the ideal and full nonlinear mode-matching for both transversal axes between the laser and the MPC was not possible with the available curved mirrors in the lab by the time of the experiment. Therefore, the initial pulse was chirped from 220 fs to 276 fs and the peak power was reduced from 400x to 350x higher than the critical power in FS at 1037 nm. It is worth noting that the initial asymmetric beam profile resulted in superior mode-matching in one axis compared to the other. After the MPC, the output beam is collimated with a spherical mirror (ROC = 5 m). We measured the autocorrelation (AC) trace after MPC directly (before compression) with a FWHM of 621 fs, and it fits well with the calculated AC of the output electric field of 635 fs using the 3D simulation.

Then, the beam undergoes 17 reflections on dispersive mirrors with different GDD profiles with an overall GDD of -3000 fs², in close agreement with the estimated GDD from the simulation (-2800 fs²). We characterized our compressed pulses using an SHG-FROG. The results are shown in Fig. 6. The measured and reconstructed traces in Fig. 6(a) and 6(b) agree well with each other, with a calculated FROG error of 0.32% in a 1024×1024 grid. Fig. 6(c) shows the measured spectrum after the compression stage, along with the reconstructed spectrum and spectral phase from the FROG measurement. The measured spectrum shows typical SPM features, where it is the main contribution to the nonlinearity. Variations in the modulation depth of the SPM lobes is attributed to residual high-order dispersion that varies the amount of energy in the wings for different B-integrals. The measured and the simulated spectra exhibit good consistency in term of bandwidth and FTL. Nevertheless, one notices a minor discrepancy in the spectral shape and more specifically the number of SPM oscillations (thus the accumulated nonlinear phase). We attribute this discrepancy to the initial spectral profile which deviates from a soliton shape as used for the simulations. Furthermore, deviations from perfect mode-matching are common and also result in varying B-integral. The compressed pulse intensity profile (Fig. 6(d)) is close to the FTL with a peak power of 10.2 GW, enhancing the input peak power by 6.4, with a FWHM pulse duration of 27 fs. Moreover, 78% of the pulse energy remains in the main pulse (2×FWHM). Adding more GDD resulted in pulses with more energy in the pedestals and caused a reduction of the peak power.

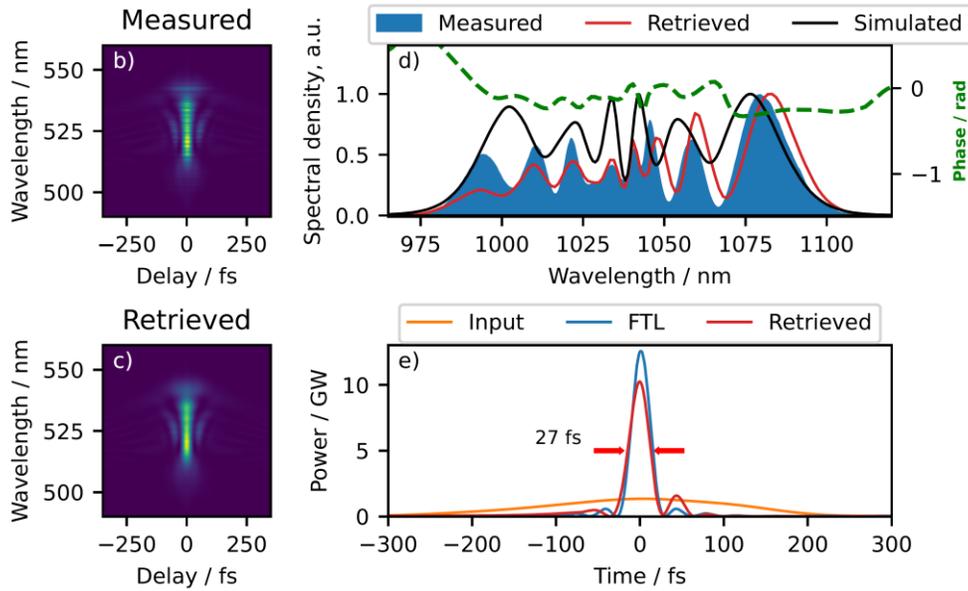

**Fig. 6.** Characterization of compressed pulse after MPC. a) Measured FROG trace. b) Retrieved FROG trace on a grid of 1024 × 1024 and an error of <0.32%. c) Measured spectral intensity with OSA and the retrieved spectrum from FROG. The simulated spectrum is depicted with a black line. The phase is displayed as a dashed green line. d) the input, the retrieved temporal pulse with FTL of the measured spectrum. The compressed pulse of 27 fs is close to the given bandwidth's transform limit (solid blue).

We investigated our compressor's spectral homogeneity, by measuring the commonly used *V*-parameter. We note however, that this is only a poor indication of spatial aberrations of the beam and a full spatio-temporal picture is required (as presented later), to evaluate the focusability of the laser. The spectral homogeneity is however an important parameter for example to validate a successful FROG retrieval. We measured 70 spectra along the *x*-axis and *y*-axis of the collimated beam. We used a multimode fiber with a mode size of 105 µm (FC/PC connector) coupled into an OSA. The fiber was positioned on an *xy*-motorized stage and fixed in z-direction for repeatable and accurate measurements with a step size of 200 µm and. Fig. (a) and 7(b) show the reconstructed spectral distributions of both axes. Fig. 7(c) and 7(d) show the *V*-parameter (solid red), and the intensity-weighted average *V*-parameter is 98.1% and 98.4% for both axes within a 14 mm diameter of the measurement area. Although the homogeneity value is excellent with respect to typical values reported in the literature, we observe a slight degradation of the *V*-parameter that is not reproduced by the simulation. Our simulation showed a minor degradation in the beam quality in this experiment in case of starting from a diffraction-limit spatial beam profile.

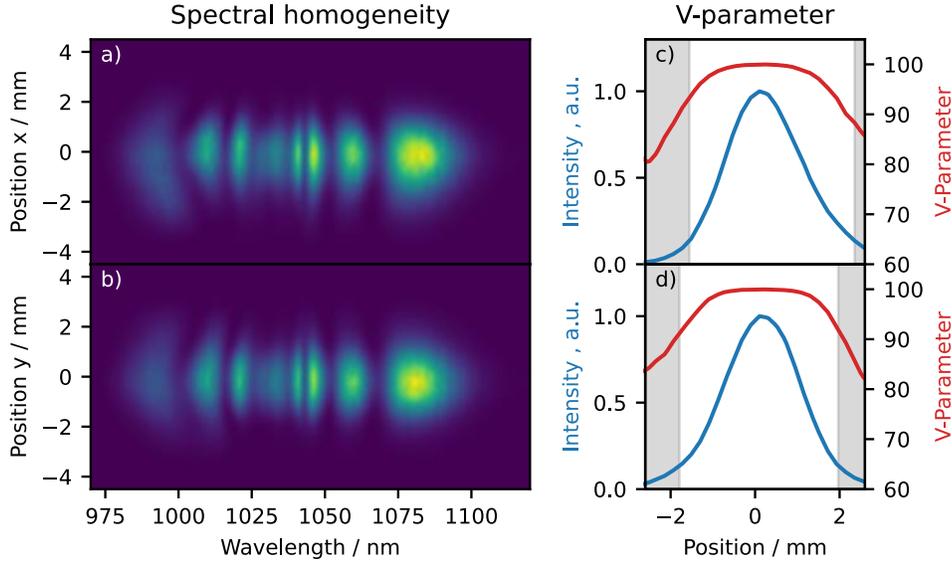

**Fig. 7.** Spatio-spectral homogeneity characterization of the compressed pulse for both axes. a) and b) maps are the reconstructed spectral distributions of the compressed pulses in $x$ and $y$ axes. c) and d) are the *V*-parameter results, where the solid blue lines on the right show the normalized intensity of the spectral profile. The red solid lines are the corresponding spatio-spectral homogeneity values (*V*-parameter), and the gray boxes show the beam area within a beam diameter ($1/e^2$).

To validate the beam quality preservation after compression, we have measured the $M^2$. Fig 8(a) shows the $M^2$ measurement of the input beam before the mode-matching setup with an $M^2$ in both axes of 1.14, with estimated waist relative astigmatism (ESA) of 28%. Fig. 8(b) shows a slight increase of the $M^2$ after propagation through the multipass compressor of 1.17×1.23, showing only a negligible beam degradation. The final ESA is measured to be 34% which is close to that measured at the input. This astigmatism is most likely caused by misalignment in the $M^2$ measurement, rather than by the MPC. This assumption is confirmed by the spatio-temporal characterization by INSIGHT, which indicates negligible astigmatism retrieved from the beam caustic.

It is well-known that spatio-temporal couplings introduce aberrations that disrupt the ideal wavefront which lead to deviations from the ideal Gaussian shape and reduced peak intensity at the focus. While the V-parameter measures the level of spectral homogeneity throughout the beam profile, it lacks the capability to convey phase information essential for evaluating the focusability of the compressed beam profile. In other words, it does not directly gauge the pulse duration within the beam profile, as crucial parameters for reaching high peak intensity in numerous applications.

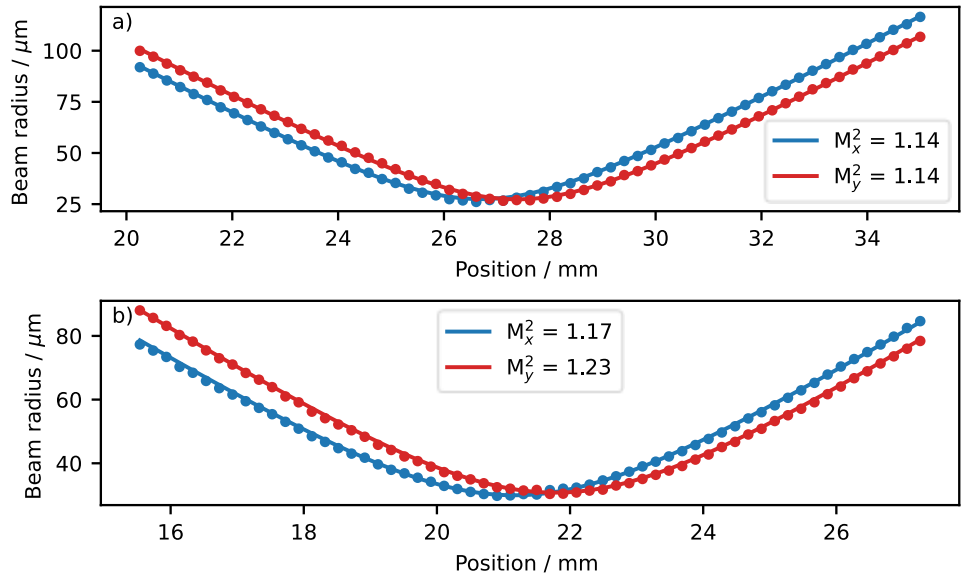

**Fig. 8.** Beam quality factor measurement ($M^2$). a) of the input beam before MPC setup. b) of the output beam along the transverse axes.

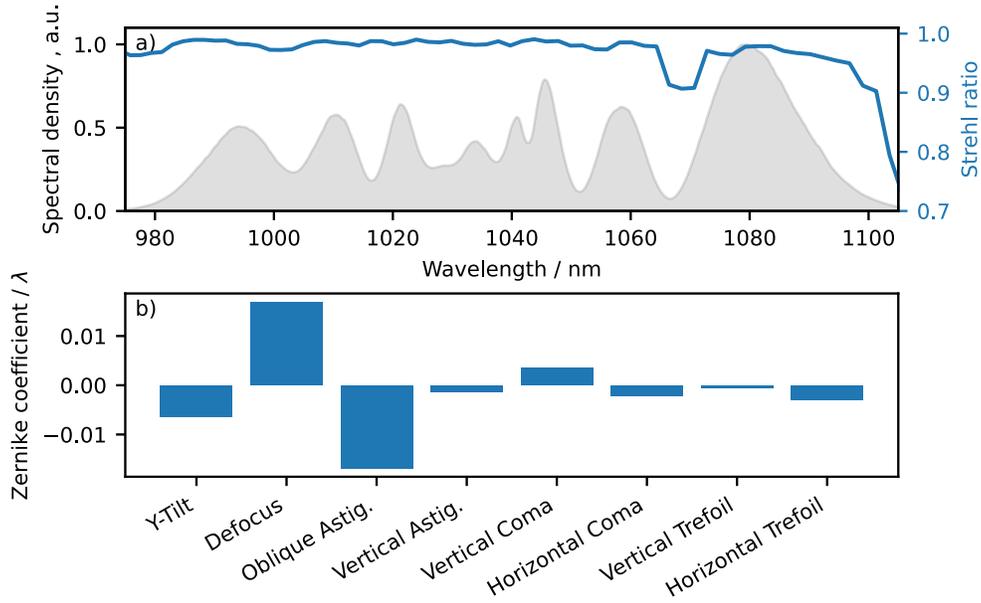

**Fig. 9.** Retrieved results of INSIGHT of the compressed beam after MPC. a). The Strehl ratio across the spectrum. b) Integrated Zernike coefficients of the geometrical aberrations such as wavefront tilt, defocus, and astigmatism.

In order to characterize the capacity of a compressor to preserve the beam quality, we fully characterized the spatio-temporal properties of the output of our compressor. In this goal, we attenuated the beam and guided it using a spherical mirror with a ROC of 1 m into the INSIGHT setup. INSIGHT employs standard Fourier-transform spectroscopy to enable the determination of the frequency-resolved spatial intensity profile across the beam, and the spatial phase profile across the beam at each frequency. INSIGHT gives directly access to spectrally-resolved 2D-maps of amplitude and phase [18–22].

One of the essential metrics to quantify the focusability of the laser beam is the Strehl ratio (SR) as it measures the efficiency of a laser beam relative to an ideal, diffraction-limited beam [23]. By determining the SR, we gain comprehension into how the aberrations affect laser beam quality and can subsequently enhance its focus. Fig. 9(a) shows the retrieved SR across the spectrum of the compressed pulse with a mean SR of 0.97. Furthermore, Zernike coefficients quantitatively measure the aforementioned aberrations, enabling wavefront and beam quality assessment. Fig. 9(b) depicts the integrated Zernike coefficients over the whole spectrum for the most common aberrations of the compressed beam after the MPC, and these coefficients are below $0.02\lambda$ where $\lambda$ is 1037 nm in this work, demonstrating that our MPC has produced a high-quality beam in the near field with minimal Zernike coefficients related to aberrations. The beam is almost diffraction-limited, meaning that it has a small divergence angle and can be focused to a small spot size. At the focus, the variation of the lateral position of the focal spot, due to frequency-dependent horizontal and vertical wavefront tilts, is less than $0.01\lambda$ of the beam waist across the laser spectrum. Oblique astigmatism and the defocus have the highest Zernike coefficient of $0.02\lambda$, which we attribute to small misalignment in the INSIGHT measurement setup.

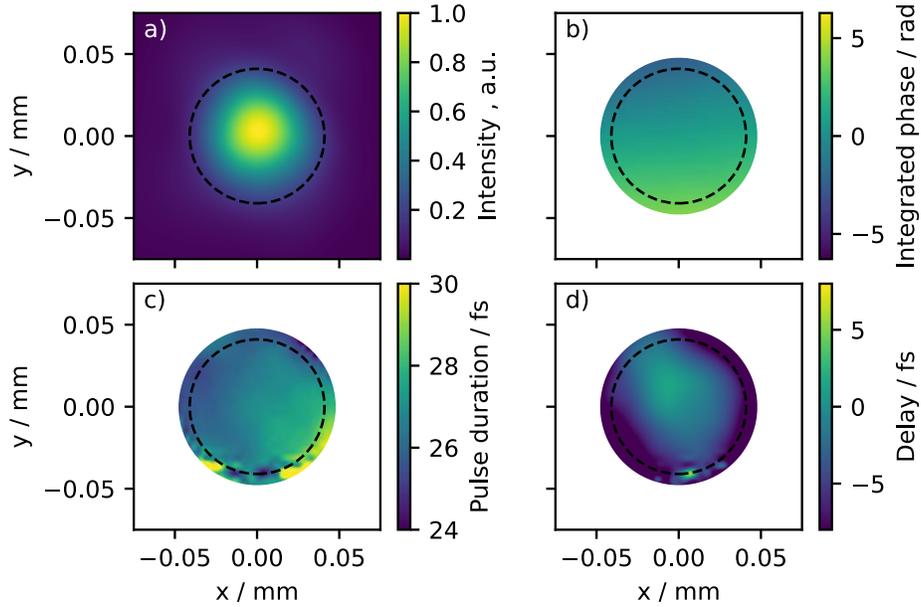

**Fig. 10.** Overview of the retrieved spatio-temporal results of INSIGHT at the focus. a) Reconstructed spectrally-integrated beam profile and the dashed circle indicates the $1/e^2$ area. b) The integrated phase across the beam profile. c) The pulse duration map across the beam profile. d) The delay of the pulse arrival at the focus across the beam profile.

Fig. 10(a) and 10(b) illustrate the reconstructed spectrally-integrated beam profile and the corresponding wavefront phase map, respectively. The root mean square (RMS) of the wavefront flatness calculated from Zernike coefficients is $0.03\lambda$, showing a quasi-unaberrated wavefront of high focusability, and it is important to note this value reaches the limit of the INSIGHT sensors. The phase map exhibits a small tilt in the $y$-direction but is negligible in the $x$-direction, resulting in a Zernike coefficient for $y$-tilt of $0.01\lambda$, and we attribute this to an oblique injection of the beam to be characterized onto INSIGHT. Importantly, this tilt does not significantly degrade the peak intensity at the focus. Fig. 10(c) showcases the temporal homogeneity across the beam profile and reveals a pulse duration of sub-28 fs within the $1/e^2$ area (black dash circle). However, a minor degradation is observed in the lower part of the circle at the edge of the $1/e^2$ region, with pulse durations still below 33 fs. This, however, does not substantially impact the spatially-integrated pulse duration, as measured by other techniques such as an AC and FROG. Fig. 10(d) depicts the temporal pulse arrival at the focus across the beam profile with root mean square of 3 fs, demonstrating that our ultrashort pulses exhibit minimal pulse front tilt and no significant chromatic phase aberration.

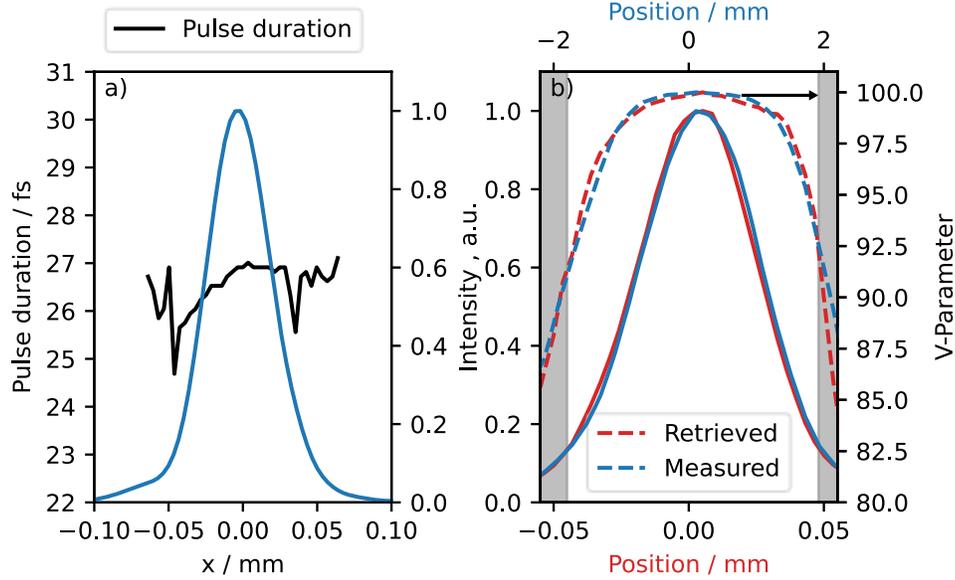

**Fig. 11.** The retrieved temporal and spectral homogeneity by INSIGHT. a) The retrieved pulse duration across the beam profile at the focus of INSIGHT. b) The retrieved and the measured V-parameter by INSIGHT and V-parameter stage of x-axis, where the solid lines (red and blue) on the right show the normalized intensity of the retrieved and measured beam profile. The dash lines are the corresponding retrieved and measured spatio-spectral homogeneity values (V-parameter), and the gray boxes show the beam area within a beam diameter ($1/e^2$).

Looking in more detail into the temporal homogeneity of the compressed pulses, Fig. 11(a) depicts the pulse duration across the beam profile at the focus. The retrieved value by INSIGHT aligns well with the measured pulse duration by the AC and the retrieved value by FROG with a mean value of 26.5 fs and RMS error of 0.7 fs across the beam within $1/e^2$ area. This temporal homogeneity is supported by the measured spatio-spectral homogeneity in Fig 7, and it aligns well with the retrieved V-parameter by INSIGHT. Moreover, in contrast to INSIGHT, the measured V-parameter is a direct spectrometer-based measurement for the spectral homogeneity, and it depends primarily on the calibration of the spectrometer. To relate both, Fig. 11(b) illustrates the retrieved -by INSIGHT- and measured V-parameter with the corresponding beam profile in both cases. The difference in the beam size managed by overlapping the $1/e^2$ of beams, where the lower x-axis represents the retrieved beam profile and the upper x-axis represent the measured beam profile. both V-parameter values show remarkable agreement within $1/e^2$, with values above 90%. One can correlate the degradation of the V-parameter beyond $1/e^2$ (beyond 0.05 mm) below 90% in Fig. 11(b) with the fluctuation of the pulse duration in Fig. 11(a). Finally, the consistency of the retrieved and measured spectral-homogeneity parameters further validate the INSIGHT results. In principle this seems to indicate that the V-parameter is thus a good, simplified metric to evaluate spatio-temporal degradations; however, a more detailed experimental and simulation study is required to evaluate the sensitivity of the V-parameter as compared to the full-spatio-temporal characterization in case of stronger degradation. This will be addressed in a future detailed study.

These results indicate that this MPC does not generate any strong aberration and achieves the highest peak intensity. Therefore, there is no need for further enhancement of the MPC setup or utilizing any spatial filtering after the compression setup to improve the beam and the homogeneity of the beam properties across the spectrum.

## 5. Conclusion

In this study, we propose a robust and cost-effective approach of nonlinear pulse compression at high pulse energy up to 400 µJ by utilizing a hybrid bulk-air MPC, eliminating the need for engineered GDD and pressure-controlled cells, and employing standard HR optics instead. Guided by 3D simulations, we achieved exceptional MPC performance by striking a balance between the nonlinear properties of ambient air and fused silica. Using our simple, single-stage method, we compressed the pulses of a commercially available, 40 W, Yb-based laser operating at 100 kHz from an initial duration of 220 fs down to 27 fs, marking a significant 8x reduction. Notably, this compression translates into a remarkable peak power enhancement from 1.6 GW to 10.2 GW with 78% of the pulse energy deposited within the main pulse while maintaining a remarkable overall optical efficiency of 91%. Importantly, spectral homogeneity remains uncompromised, with an average spectral overlap of >98% across the beam area. INSIGHT measurements revealed that the compressed pulses exhibit outstanding spectral and temporal homogeneity, minimal spatio-temporal couplings, and excellent focusability, as evidenced by a mean SR of 0.97 and low Zernike coefficients. To our knowledge, this work is the first demonstration of a hybrid air-bulk multipass cell within a single-stage MPC configuration, merging the strengths of ambient air and bulk material and paving the way for compact and robust gigawatt-class MPC compressors. Assisted by the 3D simulation, we aim to extend this cost-effective approach to higher pulse energies and few-cycle regime.

**Disclosures**

The authors declare no conflicts of interest.

**Funding:**

Ruhr-Universität Bochum (Open Access Publication Fund). Mercator Research Center Ruhr. Deutsche Forschungsgemeinschaft (DFG, German Research Foundation) under Germanys Excellence Strategy‐EXC-2033‐Projektnummer 390677874 - RESOLV. European Research Council (ERC) under the European Union's Horizon 2020 research and innovation programme (grant agreement No. 805202 - Project Teraqua)

**Acknowledgments**

We acknowledge support by the DFG Open Access Publication Funds of the Ruhr-Universität Bochum. These results are funded by the Deutsche Forschungsgemeinschaft (DFG, German Research Foundation) under Germanys Excellence Strategy‐EXC-2033‐Projektnummer 390677874 - RESOLV. This research is part of a project that has received funding from the European Research Council (ERC) under the European Union's Horizon 2020 research and innovation programme (grant agreement No. 805202 - Project Teraqua). This publication was funded by Mercator Research Center Ruhr GmbH within the project "Tailored fs-XUV Beamline for Photoemission Spectroscopy."

We thank Prof. Dr. Martina Havenith-Newen and Dr. Claudius Hoberg (Center for Solvation Science ZEMOS at Ruhr-Universität Bochum) for their help and fruitful discussions.

**Data availability**

Data underlying the results presented in this paper are not publicly available at this time but may be obtained from the authors upon reasonable request.